\documentstyle[12pt,a4]{article}

\begin{document}
\sloppy
\thispagestyle{empty}

\hfill\mbox{NIKHEF 95-051}
\vspace*{\fill}
\begin{center}
{\LARGE\bf Twist-three distributions and their}

\vspace{2mm}
{\LARGE\bf appearance in the doubly-polarized} \\

\vspace{2mm}
{\LARGE\bf Drell-Yan process%
\footnote{Talk presented  by R.D. Tangerman at 2nd Meeting on Possible
Measurements of Singly Polarized pp and pn Collisions at HERA
(DESY-IfH Zeuthen, August 31 - September 2, 1995).}}\\

\vspace{2em}
\large
R.D. Tangerman and P.J. Mulders
\\
\vspace{2em}
{\it  National Institute for Nuclear Physics and High Energy Physics}
 \\{\it  (NIKHEF), P.O. Box 41882, 1009~DB Amsterdam, The Netherlands}\\
\end{center}
\vspace*{\fill}
\begin{abstract}
\noindent
The twist-three distributions $g_2(x)$ and $h_2(x)$ are defined as quark-field
matrix elements between polarized hadron states.
They can be written in terms of quark-mass and gluonic operators, after which
the Burkhardt-Cottingham sum rule for $g_2$ can be derived and a similar sum
rule for $h_2$. Their role in the Drell-Yan double-spin asymmetry $A_{LT}$
is explained.
\end{abstract}
\vspace*{\fill}

%
\section{Introduction}
\label{sect1}

Although the basic ingredients for high-energy scattering experiments, the
quark distribution functions defined as quark-field matrix elements,
can not be calculated from QCD directly, certain sum rules have a more general
origin. The Adler sum rule, for instance, follows from current conservation.
At sub-leading twist, one has the Burkhardt-Cottingham (BC) sum rule for $g_2$,
and a similar sum rule for $h_2$, which follow from rotational
invariance~\cite{burk93,tang94}. First attempts have been made to measure the
BC
sum rule~\cite{adam94}. The measurement of the $h_2$ sum rule may be
possible in the doubly-polarized Drell-Yan process.

Within classes of equal twist, parton distribution functions can be
distinguished by their spin-content.
For unpolarized hadrons, for instance, one has the light-cone
correlator\footnote{We use the nomenclature and conventions of Jaffe and Ji in
Ref.~\cite{jaff92}.}
\begin{equation}
\int \frac{d \lambda}{4\pi}\,e^{i\lambda x}\,
\langle P|\overline{\psi}(0)\gamma^\mu\psi(\lambda n)|P\rangle
=f_1(x)\, p^\mu+f_4(x)\, M^2 n^\mu,
\end{equation}
where the hadron momentum reads $P^\mu=p^\mu+(M^2/2)n^\mu$ in terms
of the vectors $p$ and $n$, satisfying $p^2=n^2=0$ and $p\cdot n=1$.
Not an unimportant detail is the use of the light-cone gauge $A\cdot n=0$,
such that the path-ordered link operator
${\cal P}\exp[-i g\int_0^\lambda d\mu\, A(\mu n)\cdot n]$,
inserted between the quark fields to ensure color gauge invariance,
becomes unity.
The twist-two momentum distribution $f_1(x)$ will contribute to the leading
DIS structure function $F_T(x_B,Q^2)$, the logarithmic $Q^2$ dependence coming
from radiative corrections (GLAP evolution).

If the hadron is polarized with spin vector
$S^\mu=(S\cdot n)p^\mu+S_T^\mu-(M^2/2)(S\cdot n)n^\mu$, the axial vector
matrix element is parametrized like
\begin{equation}
\int \frac{d \lambda}{4\pi}\,e^{i\lambda x}\,
\langle PS|\overline{\psi}(0)\gamma^\mu\gamma_5\psi(\lambda n)|PS\rangle
=g_1(x)\, (S\cdot n) p^\mu+g_T(x)\, S_T^\mu+g_3(x)\, M^2(S\cdot n) n^\mu,
\end{equation}
where $g_1(x)$ is the twist-two helicity distribution. In this note we will
concentrate on the combination $g_2(x)=g_T(x)-g_1(x)$, containing the
transverse spin distribution, and its longitudinal counterpart
$h_2(x)/2=h_L(x)-h_1(x)$ defined by the light-cone correlator
\begin{eqnarray}
\lefteqn{\int \frac{d \lambda}{4\pi}\,e^{i\lambda x}\,
\langle PS|\overline{\psi}(0)i\sigma^{\mu\nu}\gamma_5
\psi(\lambda n)|PS\rangle=}&&\nonumber\\
&&\frac{h_1(x)}{M}\,(S_T^\mu p^\nu-S_T^\nu p^\mu)
+h_L(x)\, M (S\cdot n)(p^\mu n^\nu-p^\nu n^\mu)
+h_3(x)\, M (S_T^\mu n^\nu-S_T^\nu n^\mu),
\end{eqnarray}
where $h_1(x)$ is the twist-two transversity distribution.

\section{Sum rules}
\label{sect2}

The parton interpretation of the higher-twist distribution function is much
more
involved than for the twist-two case. This arises because of the mixing in the
operator-product-expansion of gluonic and quark-mass
operators~\cite{jaff90,jaff91}. Consider first the transverse-spin distribution
$g_2(x)$. In Ref.~\cite{tang94} it is shown to be closely related to the
existence of intrinsic quark transverse momentum in the nucleon.
Using the Dirac equation and Lorentz covariance, it can be rewritten
as
\begin{equation}
g_2(x)=-g_1(x)+\int_x^1\frac{d y}{y}g_1(y)
+\frac{m}{M}\left[\frac{h_1(x)}{x}-\int_x^1\frac{d y}{y}
\frac{h_1(y)}{y}\right]
+\tilde{g}_T(x)-\int_x^1\frac{d y}{y}\tilde{g}_T(y),\label{aap}
\end{equation}
where $m$ is the current-quark mass.
The distribution $\tilde{g}_T$ contains explicit gluon fields, and is defined
by
\begin{equation}
x\tilde{g}_T(x)\, S_T^\mu
=\int\frac{d\lambda}{8\pi}\,e^{i\lambda x}\,
\langle PS|\overline{\psi}(0)gA_{T\nu}(0)[g_T^{\mu\nu}\not\!{n}\gamma_5
+i\epsilon_T^{\mu\nu}\not\! n]\psi(\lambda n)|P\rangle+\mbox{h.c.},
\end{equation}
using the transverse tensors $g_T^{\mu\nu}=g^{\mu\nu}-p^\mu n^\nu-p^\nu n^\mu$
and $\epsilon_T^{\mu\nu}=\epsilon^{\mu\nu\rho\sigma}p_\rho n_\sigma$.
The decomposition~(\ref{aap}) is interesting, since it gives rise to the
Burkhardt-Cottingham sum rule~\cite{burk70}
\begin{equation}
\int_0^1 dx\,g_2(x)=0,
\end{equation}
provided the order of $x$ and $y$ integrations may be interchanged.
This sum rule has received some attention recently. Specifically,
it was shown that for a free-quark target it acquires no
${\cal O}(\alpha)$ corrections~\cite{alta94}.

In a similar fashion, the longitudinal-spin distribution $h_2(x)$ can be
decomposed as
\begin{eqnarray}
\frac{h_2(x)}{2}&=&-h_1(x)+2x\int_x^1\frac{d y}{y^2}h_1(y)
+\frac{m}{M}\left[\frac{g_1(x)}{x}-2x\int_x^1\frac{d y}{y^2}
\frac{g_1(y)}{y}\right]\nonumber\\
&&+\tilde{h}_L(x)-2x\int_x^1\frac{d y}{y^2}\tilde{h}_L(y),\label{noot}
\end{eqnarray}
with the interaction-dependent function
\begin{equation}
x\tilde{h}_L(x)\, M (S\cdot n)
=\int\frac{d\lambda}{8\pi}\,e^{i\lambda x}
\langle PS|\overline{\psi}(0)g\!\!\not\! {A_T}(0)\,\gamma_5\!\!\not\!{n}
\psi(\lambda n)|PS\rangle+\mbox{h.c.}.
\end{equation}
Equation~(\ref{noot}) leads to the sum rule~\cite{burk93,tang94}
\begin{equation}
\int_0^1 dx\,h_2(x)=0,
\end{equation}
for the {\em distribution\/} function $h_2$. The validity of this sum rule for
the {\em structure\/} function $h_2$ was questioned recently by
Burkardt~\cite{burk93,burk95}.

\section{Experiments}
\label{sect3}

Writing down sum rules for parton distribution functions is one thing,
extracting them from experiment is another. The distribution $g_2(x)$ appears
cleanest in the cross section for the scattering of a longitudinally polarized
lepton on a transversely polarized hadron, measuring only the scattered lepton
momentum. From the handbag diagram in Fig.~1 and two diagrams involving a
quark-gluon-quark correlator, one finds that the
structure function $g_2(x_B,Q^2)$ at tree-level equals
$(1/2)\sum_a e_a^2\, g_2^a(x_B)$, the $a$ denoting quark and antiquark flavors.
Since $h_2(x)$ is chirally odd, one needs a {\em second\/} hadron to flip
helicities. Thus we are forced to consider, for instance, Drell-Yan scattering.
The analogue to the handbag diagram is depicted in Fig.~2.
After averaging over the transverse momentum $q_T$ of the produced massive
photon, the distributions appear pair-wise in the double-spin
asymmetry~\cite{jaff92,tang94}
\begin{eqnarray}
A_{LT}=&&\frac{\sigma(\lambda_A,S_{BT})-\sigma(\lambda_A,-S_{BT})}%
{\sigma(\lambda_A,S_{BT})+\sigma(\lambda_A,-S_{BT})}
=\frac{M\lambda_A}{Q}\frac{\sin 2\theta\cos(\phi-\phi_B)}{1+\cos^2\theta}
\nonumber\\&&\times
\frac{\sum_a e_a^2 \left\{g_1^a(x) y\left[g_T^{\bar{a}}(y)
+\tilde{g}_T^{\bar{a}}(y)\right]+x\left[h_L^a(x)
+\tilde{h}_L^a(x)\right]h_1^{\bar{a}}(y)\right\}}{\sum_a e_a^2 f_1^a(x)
f_1^{\bar{a}}(y)}.
\end{eqnarray}
The helicity of the longitudinally polarized hadron is given by
$\lambda_A=S_A\cdot n$. The angles $\theta$ and $\phi$ are of those the
lepton axis in the photon rest-frame.
Also, $x=P_B\cdot q/P_A\cdot P_B$ and $y=P_A\cdot q/P_A\cdot P_B$.
It is clear that $h_2(x)$ cannot easily be extracted from the above
expression, since it pre-supposes knowledge of the other distributions.
Also, the combinations $g_T+\tilde{g}_T$ and $h_L+\tilde{h}_L$
appear, making life even more complicated. Simply neglecting quark
transverse momentum such that (for zero quark mass) $\tilde{g}_T=g_T$
and $\tilde{h}_L=h_L$, as is done in Ref.~\cite{jaff92}, is an
inconsistent approximation~\cite{tang94}.
\\[1cm]
This work is part of the research program of the foundation for
Fundamental Research of Matter (FOM) and the National Organization
for Scientific Research (NWO).

\newpage



\input feynman
\bigphotons

\begin{figure}[hb]
\begin{picture}(39000,22000)(0,-1000)

\put(0,2000){
\begin{picture}(20000,18000)(-8000,-9000)

\drawline\fermion[\N\REG](-4000,-4000)[4400]
\drawarrow[\LDIR\ATBASE](\pmidx,\pmidy)
\drawline\photon[\NW\CURLY](\pbackx,\pbacky)[4]
\put(\pbackx,\pbacky){\makebox(0,0)[br]{$q\ $}}
\global\advance\pmidy by 1000
\put(\pmidx,\pmidy){\vector(1,-1){1200}}
\drawline\fermion[\E\REG](\pfrontx,\pfronty)[8000]
\global\advance\pmidx by 500
\drawarrow[\LDIR\ATBASE](\pmidx,\pmidy)
\drawline\photon[\NE\CURLY](\pbackx,\pbacky)[4]
\put(\pbackx,\pbacky){\makebox(0,0)[bl]{$\ q$}}
\global\advance\pmidy by -1000
\put(\pmidx,\pmidy){\vector(1,1){1200}}
\drawline\fermion[\S\REG](\pfrontx,\pfronty)[4400]
\drawarrow[\LDIR\ATTIP](\pmidx,\pmidy)
\put(-4300,-3500){\vector(0,1){750}}
\put(4300,-2750){\vector(0,-1){750}}
\put(-3800,-3500){\makebox(0,0)[bl]{$k$}}
\put(3800,-3500){\makebox(0,0)[br]{$k$}}
\multiput(0,-600)(0,400){5}{\line(0,1){200}}

\put(0,-5000){\oval(10000,2000)}
\multiput(0,-6500)(0,400){8}{\line(0,1){200}}
\linethickness{0.1cm}
\drawline\fermion[\W\REG](-5000,-5000)[2000]
\drawline\fermion[\E\REG](5000,-5000)[2000]

\end{picture}}
\put(2000,1000){\parbox{3in}{Fig.~1. Handbag diagram.}}

\put(20000,2000){
\begin{picture}(20000,18000)(-10000,-10000)

\drawline\fermion[\N\REG](-4000,-5000)[5000]
\drawarrow[\LDIR\ATBASE](\pmidx,\pmidy)
\drawline\photon[\E\REG](\pbackx,\pbacky)[2]
\global\advance\pmidy by 500
\put(\pmidx,\pmidy){\vector(1,0){750}}
\global\advance\pmidy by -1000
\put(\pmidx,\pmidy){\makebox(0,0)[t]{$q$}}
\drawline\fermion[\N\REG](\pfrontx,\pfronty)[5000]
\drawarrow[\LDIR\ATBASE](\pmidx,\pmidy)
\drawline\fermion[\N\REG](4000,-5000)[5000]
\drawarrow[\S\ATTIP](\pmidx,\pmidy)
\drawline\photon[\W\FLIPPED](\pbackx,\pbacky)[2]
\global\advance\pmidy by 500
\global\advance\pmidx by -750
\put(\pmidx,\pmidy){\vector(1,0){750}}
\global\advance\pmidy by -1000
\put(\pmidx,\pmidy){\makebox(0,0)[t]{$q$}}
\drawline\fermion[\N\REG](\pfrontx,\pfronty)[5000]
\drawarrow[\S\ATTIP](\pmidx,\pmidy)
\put(-4300,-4500){\vector(0,1){750}}
\put(4300,-3750){\vector(0,-1){750}}
\put(-3800,-4500){\makebox(0,0)[bl]{$k_a$}}
\put(3800,-4500){\makebox(0,0)[br]{$k_a$}}
\put(-4300,4500){\vector(0,-1){750}}
\put(4300,3750){\vector(0,1){750}}
\put(-3800,4500){\makebox(0,0)[tl]{$k_b$}}
\put(3800,4500){\makebox(0,0)[tr]{$k_b$}}

\put(0,-6000){\oval(10000,2000)}
\multiput(0,-7600)(0,400){8}{\line(0,1){200}}
\linethickness{0.1cm}
\drawline\fermion[\W\REG](-5000,-6000)[2000]
\drawline\fermion[\E\REG](5000,-6000)[2000]
\thinlines

\put(0,6000){\oval(10000,2000)}
\multiput(0,7600)(0,-400){8}{\line(0,-1){200}}
\linethickness{0.1cm}
\drawline\fermion[\W\REG](-5000,6000)[2000]
\drawline\fermion[\E\REG](5000,6000)[2000]
\thinlines

\end{picture}}

\put(20000,1000){\parbox{3in}{Fig.~2. Drell-Yan tree-level diagram.}}
\end{picture}
\end{figure}

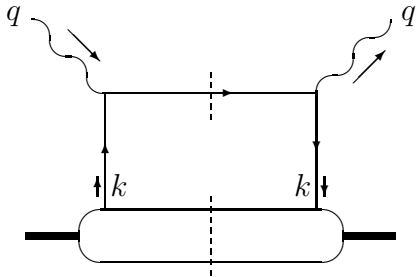
\end{document}